\documentclass[prb,superscriptaddress,twocolumn,preprintnumbers,a4,showpacs,floatfix]{revtex4}  
\usepackage{amsmath, amssymb, psfrag, tabls, dcolumn, bm, color}
\usepackage[sort&compress]{natbib}
\usepackage{graphicx}
\usepackage{textcomp}

\newcommand{\be}{\begin{equation}} 
\newcommand{\ee}{\end{equation}}
\newcommand{\bma}{\begin{displaymath}}
\newcommand{\ema}{\end{displaymath}}

\begin{document} 

\title{Gold in graphene: in-plane adsorption and diffusion}

\author{Sami Malola}
\address{Department of Physics, NanoScience Center, 40014 University of Jyv\"askyl\"a, Finland}

\author{Hannu H\"akkinen}
\address{Department of Physics, NanoScience Center, 40014 University of Jyv\"askyl\"a, Finland}
\address{Department of Chemistry, NanoScience Center, 40014 University of Jyv\"askyl\"a, Finland}

\author{Pekka Koskinen\footnote{Author to whom correspondence should be addressed.}}
\email{pekka.j.koskinen@jyu.fi}
\address{Department of Physics, NanoScience Center, 40014 University of Jyv\"askyl\"a, Finland}


\date{\today}

\begin{abstract} 
We study the bonding and diffusion of Au in graphene vacancies using density-functional theory. Energetics show that Au adsorbs preferably to double vacancies, steadily in-plane with graphene. All diffusion barriers for the complex of Au in double vacancy are above $4$~eV, whereas the barriers for larger vacancies are below $2$~eV. Our results support the main results of a recent experiment [Gan \emph{et al.}, \emph{Small} {\bf 4}, 587 (2008)], but suggest that the observed diffusion mechanism is not thermally activated, but radiation-enhanced.
\end{abstract}

\maketitle

%
%
As the exceptional properties of graphite have been long known, it is no surprise that the recent developments in graphene fabrication techniques have given an enormous impetus for graphene research.~\cite{novoselov_science_04,geim_nmat_07} Not only is pure graphene interesting, but also interactions with metal atoms, because of their fundamental relevance to applications like catalysis, batteries, and nanoelectronics.~\cite{nnano_2_605,small_2_182}

These interactions have been investigated both experimentally and theoretically. Previous studies include metal atom adsorption and diffusion on surfaces and edges of graphite, graphene and carbon nanotubes.\cite{MtrlSciAndEC_26_1207} Among experiments, transmission electron microscopy (TEM) is a versatile method for sample characterization as well as manipulation~\cite{krasheninnikov_nmat_07,banhart_small_05,gan_NJP_08}, and can achieve nearly single atom accuracy. Recent interesting experiment by Gan, Sun and Banhart~\cite{gan_small_08}, observed in-plane adsorption of Au and Pt atoms in graphene sheets and carbon nanotubes using TEM, and also measured the rate for in-plane metal atom diffusion.

In this Letter we confirm this experimentally observed propensity of Au to adsorb in-plane with graphene. The experiment suggested thermal in-plane diffusion of Au with energy barrier $\approx2.5$~eV; our calculations suggest that the diffusion mechanism is not thermal, but radiation-enhanced.

\begin{figure}
\includegraphics[width=\columnwidth]{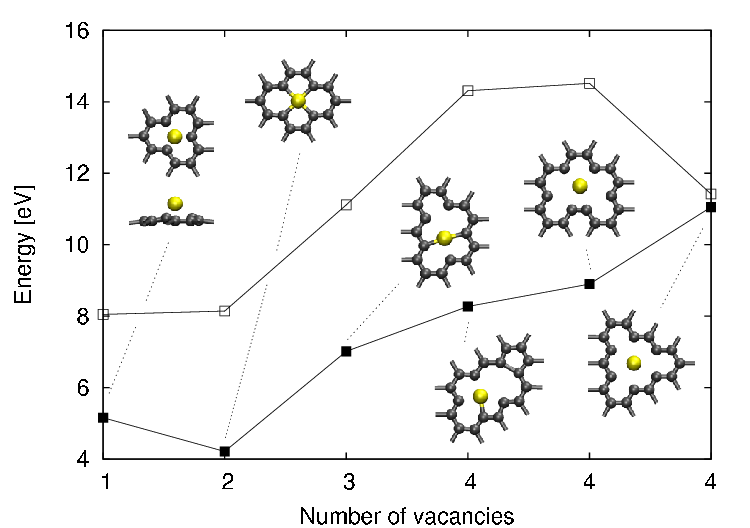}
\caption{Formation energies for carbon vacancies in graphene (empty squares), and formation energies for Au adsorbed in graphene vacancies (filled squares; local geometries shown in insets). The difference between the two curves is the Au adsorption energy. Apart from Au in single vacancy, all structures are planar.}
\label{fig1} 
\end{figure}

%
%
We used density-functional theory (DFT) with PBE functional~\cite{perdew_PRL_96}, and projector augmented waves for treating the valence electrons~\cite{blochl_PRB_94}, as implemented in real-space grid code GPAW~\cite{mortensen_PRB_05,GPAW-specs}. We started by calculating vacancy formation energies, defined for $n$-atom vacancy as
\[
E(\textrm{form})=[E(n\textrm{-vacancy})+n\cdot \varepsilon_{coh}]-E(\textrm{graphene}),
\]
for single to quadruple vacancies; here $\varepsilon_{coh}=7.9$~eV is the graphene cohesion energy.\cite{koskinen_PRL_08} Definition above is the cost to create an $n$-atom vacancy and consider the removed carbons as a part of graphene elsewhere. This definition is relevant because it appeared that carbon atoms on graphene diffuse rapidly to anneal other vacancies.~\cite{gan_small_08} The vacancy formation energies, shown in Fig.~\ref{fig1} for optimized geometries~\cite{bitzek_PRL_06}, agree with previous calculations.~\cite{PRB_74_245420} For single and double vacancies they are close to $8$~eV, and increase ultimately with $\sim2$~eV/C-atom slope for larger vacancies. Note that once a single vacancy exists, double vacancy costs practically no energy---in nanotubes double vacancy costs actually less than the single vacancy.~\cite{krasheninnikov_CPL_06} The last quadruple vacancy has exceptionally low energy due to the absence of dangling bonds in the symmetric hole.

Fig.~\ref{fig1} shows also the formation energies for Au adsorbed in graphene vancancies, where the reference is graphene and Au in vacuum. The difference between these two curves, which is the Au adsorption energy, is $3-6$~eV; last quadruple vacancy is a clear exception. Single vacancy is too small for Au and optimized geometry is not planar, whereas in larger vacancies Au is precisely in-plane with graphene. Hence, assuming a planar adsorption geometry---as demonstrated experimentally---and thermal equilibrium, we conclude that \emph{Au preferably adsorbs in double vacancies in-plane with graphene}.

%
%

\begin{figure}
\includegraphics[width=\columnwidth]{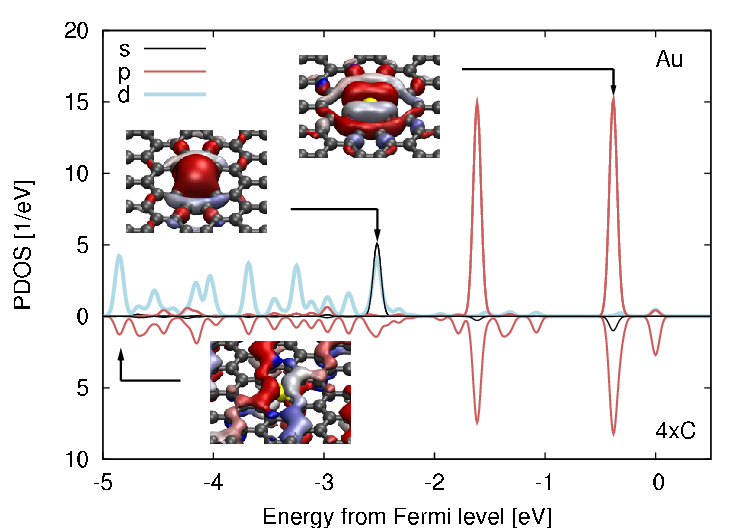}
\caption{Projected density of states for $s$-, $p$- and $d$-orbitals of Au (above abscissa) and the four neighboring C atoms (below abscissa). Insets show the wavefunctions, color standing for the phase, for selected $s$- $p$- and $d$-dominated contributions for Au. Charging of Au according to Bader analysis is $+0.5e$.
}
\label{fig2}
\end{figure}

Why does Au prefer bonding in-plane? This is analyzed chemically in Fig.~\ref{fig2}. Firstly, the in-plane orbitals of Au overlap with the dangling bonds of neighboring C-atoms in a bonding configuration; the two $p$-orbital contributions are split due to elongated structure of the vacancy. Secondly, and most importantly, out-of-plane Au $d$-orbitals bind above and below to the $\pi$-electrons of graphene---this gives Au atom stability against out-of-plane motion. The in-plane stability was confirmed also by molecular dynamics simulations at high temperatures. Because the bonding is strong both in-plane and out-of-plane, we expect the motion of Au to be confined by large energy barriers; in the following we shall investigate Au diffusion barriers in a double vacancy in more detail.

%
%
\begin{figure}
\includegraphics[height=8cm]{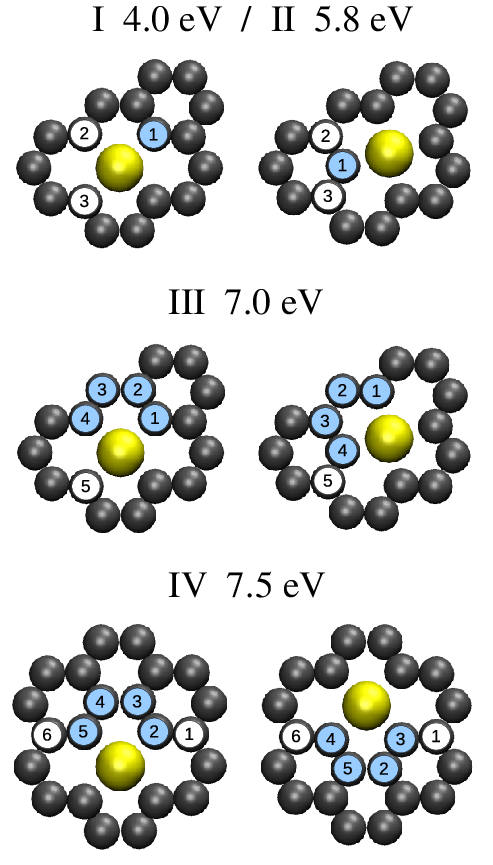}
\caption{Diffusion mechanism paths and their transition state energies for Au in double vacancy. Numbers help to visualize C atom identification during Au jump; gray (blue) for C atoms that change positions. Paths were optimized with nudged elastic band method~\cite{henkelman_JCP_00,NEB-specs} starting from almost linearly interpolated initial guess. Path I contains out-of-plane motion of Au, whereas path II contains out-of-plane motion of C(\#$1$). Path IV contains only translation, other paths also rotation.}
\label{fig3}
\end{figure}

To explore different Au diffusion paths, we used simple guessing, we conducted extensive molecular dynamics simulations with density-functional based tight-binding~\cite{koskinen_NJP_06, malola_PRB_08}, and we used constrained dynamics, for subsequent analysis. Several paths were optimized, and the most prominent ones are shown in Fig.~\ref{fig3}. The lowest-energy path I with $4.0$~eV barrier involves radical out-of-plane motion of Au, and path II with $5.8$~eV barrier involves radical out-of-plane motion of C. The lowest nearly in-plane path III has much higher barrier of $7.0$~eV, path IV with barrier even higher.

We cannot absolutely exclude the existence of other interesting paths, but we are confident that there are no relevant paths with $\approx2.5$~eV energy barriers, as suggested by the experiment.\cite{gan_small_08} Firstly, radical out-of-plane motion is unlikely because measured diffusion was quantitatively similar for Au in single graphene sheet as well as between multiple graphene sheets. Since the barrier is the same as Au adsorption energy, overcoming the barrier of path I, for example, could equally well result in the escape of Au. Secondly, any simple site-exchange mechanism or out-of-plane motion of Au is particularly unlikely because it would require large local momentum imbalance due to large $m_\textrm{Au}/m_\textrm{C}=16.4$ mass ratio. Thirdly, the lowest of the nearly in-plane paths has $7.0$~eV barrier---more paths than shown were examined. Since the in-plane motion of Au requires breaking several bonds, and since graphene is exceptionally stiff material, nearly three-fold reduction in energy barrier, for some unforeseen path, is not plausible.

On the other hand, the barrier for a single vacancy that merges with double vacancy containing Au is only $0.2$~eV. Since the triple vacancy, as shown in Fig.~\ref{fig1}, has extra space around Au, in-plane motion is easier. And indeed, it was not difficult to guess a strictly in-plane diffusion path for Au with barrier below $2$~eV. Hence, in triple vacancies Au could diffuse easily; but Au existing in triple vacancy cannot be justified on thermodynamic grounds, and, further, the barrier is too low and diffusion would have been too fast for TEM observation in the experiment of Ref.~\onlinecite{gan_small_08}.

Now we shall juxtapose our calculations with the experiment by analysing information given by Gan~\emph{et al.}~\cite{gan_small_08} To begin with, we point that the barrier $2.5$~eV was not deduced from Arrhenius-type behavior, but from the equation
\[
 D=ga^2\nu_0 \exp{(-E_b/k_BT)},
\]
where $D$ is the diffusion constant, measured by directly monitoring atoms, $g\sim1$ is a geometrical factor, $a$ is the lattice constant and $\nu_0$ is the attempt frequency. The measured diffusion constant for Au at $T=600$~\textdegree C was $D=6 \times 10^{-22}-2\times 10^{-21}$~m$^2$s$^{-1}$. Taking $a=1-2$~\AA (depends on mechanism) and $\nu_0\approx 5\times 10^{12}$~s$^{-1}$,~\cite{koskinen_PRL_08} the time between Au jumps for two-dimensional diffusion is $\sim 2-20$~s. 

How about radiation? With the carbon atom displacement threshold of $15$~eV, the radiation from $30$~A/cm$^2$ current density was estimated to cause displacement of every carbon atom in $180$~s, where most of the vacancies are rapidly annealed.\cite{banhart_RPP_99,krasheninnikov_nmat_07} Hence, from selected $N$ atoms at least one is displaced every $180/N$~s. Assuming Au in double vacancy to have $14$ neighboring C atoms, at least one of these is displaced every $\sim10$~s. Vacancy's environment is energetically excited at this rate, and the excitations may help Au to overcome large energy barriers. This rate, obtained from radiation effects, is too close to independently obtained Au diffusion jump rate to be a mere coincidence. Our conclusion, based on our calculated energetics, barrier heights, and the experimental analysis, is that the diffusion in the experiment is \emph{radiation-enhanced diffusion}.\cite{banhart_RPP_99}

The radiation-enhanced interpretation would explain some peculiar features in the experiment. First, the diffusion constant for Au and Pt was similar, which was not expected on the account that C-Pt interaction is stronger than C-Au.\cite{maiti_CPL_04} This can be now understood: if the mechanism is dominated by radiation enhancement, the C-metal interaction will not play a crucial role. Second, one-dimensional diffusion, and diffusion in the curved layers of carbon nanotubes, was very similar to pure two-dimensional diffusion. This cannot be understood if the diffusion is thermal---it is not plausible that different chemical interactions give the same barriers for Au and Pt in different environments. Using radiation-enhanced argument, in turn, carbon environment around the metal can be thought to be excited occasionally into extremely vigorous motion, causing insensitivity to interactions and geometries. We shall continue our calculations with Pt and other transition metals.

What could the actual diffusion mechanism be? The mechanism could be diffusion of Au in double vacancy. Radiation can frequently provide the required $4$~eV into the neighborhood of Au---even though every push certainly does not cause a jump. The mechanism could be merging of a single vacancy with Au in double vacancy, followed by diffusion of Au in triple vacancy, and finally followed by annealing of triple vacancy with diffusing C atoms. Complex mechanism like this would explain the temperature-dependence that was observed in Fig.~4 of Ref.~\onlinecite{gan_small_08}. There are several possibilities for the mechanism, once it is radiation-empowered.

In conclusion, because the problem has technological relevance, and as the ultimate solution is hard to provide by further calculations, the status calls for experiments to clarify the relations between thermal and radiation-enhanced diffusion.

We acknowledge support from the Academy of Finland (projects 121701 and 117997 and the FINNANO program, consortium MEP) and from the Finnish Cultural Foundation (S.M.). Computational resources were provided by the Finnish IT Center for Science (CSC).



\begin{thebibliography}{23}
\bibitem[{NEB()}]{NEB-specs}
\bibinfo{note}{The paths in Fig.~\ref{fig3} were optimized to maximum force of
  {I} \& {II}) $0.1$~{eV/\AA} (7 images in total) {III}) $0.3$~{eV/\AA} (12
  images), and {IV}) $0.3$~{eV/\AA} (11 images)}.
\end{thebibliography}

\end{document}